\documentclass[aps,twocolumn,amsmath]{revtex4}

\usepackage{graphicx}
\usepackage{amsmath}

\newcommand{\be}{\begin{equation}}
\newcommand{\ee}{\end{equation}}
\newcommand{\ba}{\begin{eqnarray}}
\newcommand{\ea}{\end{eqnarray}}

\begin{document}

\title{Structure and equation of state of interaction
site models \\ for disc-shaped lamellar colloids}

\author{Dino Costa$^1$\footnote{Corresponding author, email: %
{\tt dino.costa@unime.it}}, Jean-Pierre~Hansen$^2$, and Ludger Harnau$^{3,4}$} 
\affiliation{$^1$Dipartimento di Fisica, Universit\`a di Messina  
Contrada Papardo, C.P. 50, 98166 Messina -- Italy   \\
$^2$Department of Chemistry, University of Cambridge   
Lensfield Road, Cambridge CB2 1EW -- UK  \\
$^3$Max-Planck-Institut f\"ur Metallforschung,  
         Heisenbergstr.\ 3, D-70569 Stuttgart -- Germany \\
$^4$ Institut f\"ur Theoretische und Angewandte Physik, 
   Universit\"at Stuttgart, Pfaffenwaldring 57, D-70569 Stuttgart -- Germany}

\begin{abstract}
We apply RISM (Reference Interaction Site Model) and PRISM
(polymer-RISM)  theories to calculate the site-site pair structure
and the osmotic equation of state of suspensions of circular
or hexagonal platelets (lamellar colloids) over a range
of ratios of the particle diameter over thickness $D/\sigma$.
Despite the neglect of edge effects, the simpler PRISM theory yields
results in good agreement with the more elaborate RISM calculations,
provided the correct form factor, characterizing the intramolecular
structure of the platelets, is used.
The RISM equation of state is sensitive to the number $n$ of sites used
to model the platelets, but saturates when the hard spheres, associated 
with the interaction sites, nearly touch; the limiting equation of state
agrees reasonably well with available simulation data for all
densities up to the isotropic-nematic transition. When properly
scaled with the second virial coefficient, the equations of state of 
platelets with different aspect  ratios $D/\sigma$ nearly collapse
on a single master curve.
\end{abstract}

\maketitle

\columnseprule 0pt

\section{Introduction}
Onsager's celebrated theory~\cite{onsa:49}
of suspensions of rod-like or disc-like
cylindrical particles of diameter $D$ and length (thickness) $L$
predicts an isotropic to nematic phase transition, both in the
rod limit $L/D \gg 1$ and in the plate limit $L/D \ll 1$,
on purely entropic grounds. While Onsager's prediction received 
abundant experimental
confirmation for several lyotropic colloidal systems of elongated
rod-like particles, like the much studied TMV virus~\cite{frad:95},
the existence of such a transition in dispersions of lamellar particles,
first seen by Langmuir~\cite{lang:38}, 
was only recently confirmed by experiments on polymer-grafted gibbsite
platelets~\cite{koij:98}. 
From a theoretical point of view, there exists no equivalent,
in the case of platelets, of the Onsager limit, whereby the equation of state 
of hard rods is accurately described by the virial series truncated after second
order in the limit $L/D \to \infty$. The contribution of higher order
virial coefficients to the equation of state of the cylinders remains large, 
even in the limit $L/D \to 0$ (infinitely thin plates), except
at very low densities~\cite{eppe:84}.
For platelets of finite thickness ($L/D \ne 0$), virial
coefficients of order higher than two ($B_3$, $B_4$, $\ldots$)
are not available.
Hence there is a clear need to  develop
accurate theories for simple models of plate-like 
particles. Attempts have been made along various lines: \par 
--- Scaled particle theory~\cite{reis:59}, which is very successful for hard
sphere fluids, has been extended to prolate or
oblate hard ellipsoids of revolution~\cite{cott:79},
with moderate success when gauged against Monte Carlo 
simulations~\cite{muld:85}. Much less work has been 
devoted to disc-shaped platelets
and the results for infinitely thin platelets ($L/D=0$)~\cite{savi:81}
overestimate Monte Carlo data for the equation of state
quite significantly at densities
approaching the isotropic to nematic transition~\cite{eppe:84}.
Scaled particle theory, or its extension by Boublik~\cite{boub:75},
also overestimates the pressure of the closely related
model of ``cut hard spheres'', for which
Monte Carlo data are available for several
aspect ratios $L/D$~\cite{veer:92,zhan:02}. \par
--- Onsager theory, based on the second virial coefficient $B_2$
alone, can be ``rescaled'' according to the 
prescription of Parsons~\cite{pars:79} and Lee~\cite{slee:87}.
Although this semi-empirical procedure gives reasonably good
results for hard rod systems ($L/D > 1$), it is much
less satisfactory for hard discs ($L/D < 1$)~\cite{wens:04}. \par
--- Rosenfeld's very successful ``fundamental measure theory''
for hard sphere systems~\cite{rose:89} has been extended to the Zwanzig
model~\cite{zwan:63} for square platelets~\cite{harn:02},
but a direct comparison with disc-shaped particles is not possible because
within the Zwanzig model the platelets can only have three 
discrete orthogonal orientations. \par
\begin{figure}
\begin{center}
\includegraphics[width=5.5cm]{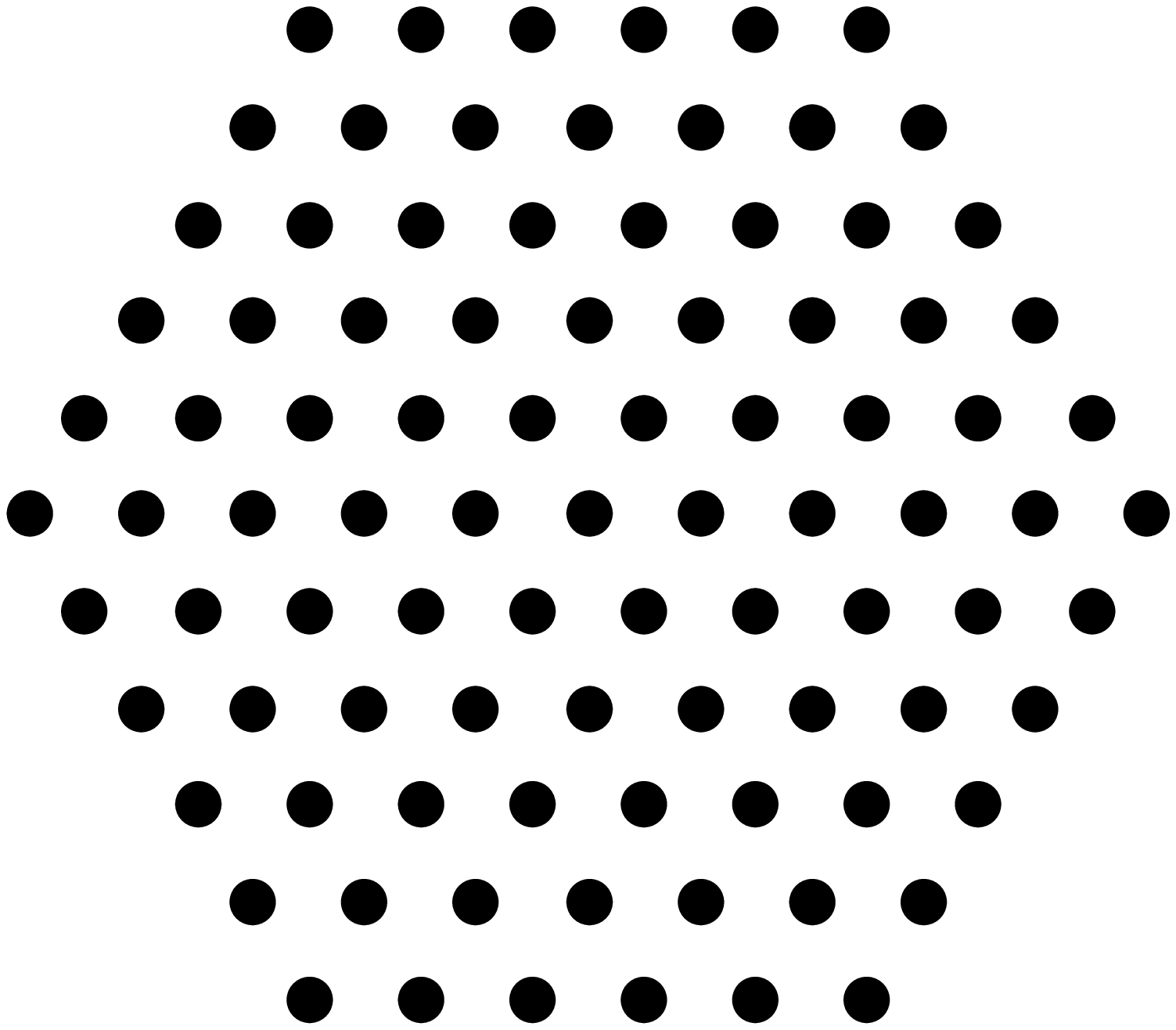}
\includegraphics[width=5.5cm]{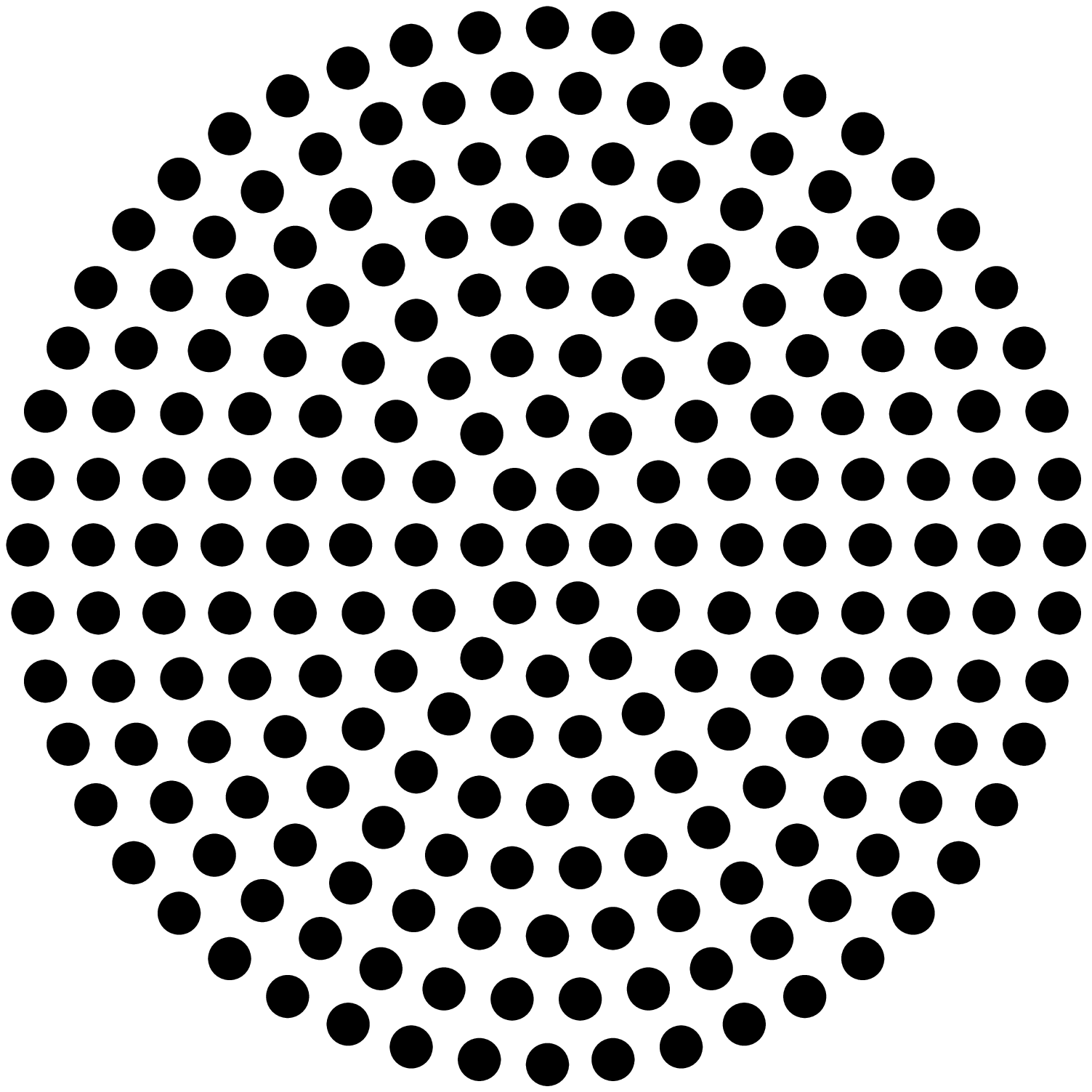}
\caption{Schematic representation of site patterns 
of model platelets
investigated in this work.
}\label{fig:model}
\end{center}
\end{figure}

--- Instead of modelling platelets as geometrically simple convex bodies
(like flat cylinders or cut spheres), an alternative is to
consider platelets to be rigid, ordered arrays of interaction sites,
distributed over the surface of a disc in some
regular pattern~\cite{kutt:00}. Two examples
considered in this paper are shown in Fig.~\ref{fig:model}.
Site-site interactions are assumed 
to be spherically symmetric, i.e. to depend only on the distance between
sites on different platelets. In Ref.~\cite{kutt:00}
the site-site pair potential was chosen to be of the screened 
Coulomb form, to mimic a dispersion of highly charged
Laponite platelets~\cite{lapo:90} in the presence of added salt,
and the pair structure was determined by Molecular Dynamics
simulations. 

In the present paper we return to this interaction site model,
but replace the screened Coulomb interaction by a hard-sphere repulsion,
in order to mimic as far as possible suspensions of uncharged,
flat, cylindrical platelets. The objective is to calculate 
the pair structure and osmotic equation of state from RISM
(Reference Interaction Site Model,~\cite{chan:72,mons:90})
and PRISM (polymer-RISM,~\cite{schw:87,schw:97})
integral equations for the site-site pair distribution
functions. A systematic comparison will be made
between the results from the RISM and PRISM
theories, in order to quantify the importance
of the pre-averaging procedure and
of ``edge'' effects, which are involved within
the simpler PRISM theory, where all sites are assumed to 
be equivalent~\cite{schw:97}. Wherever possible, the
theoretical predictions will be compared to available simulation data. 
Parts of the present work 
were briefly reported elsewhere~\cite{harn:01}.

\section{Interaction site models}

The lamellar systems under consideration contain $N$ disc-shaped
platelets of diameter $D=2R$ inside a volume $V$. Each platelet
carries a regular pattern (or grid) of $n$ interaction
sites occupied by a hard sphere of diameter $\sigma$.
We consider the two grids pictured in Fig.~\ref{fig:model}.
The circular grid is made up of $n_{\rm r}$
concentric rings of diameter 
$d_i=(i-1)(D-\sigma)/(n_{\rm r} -1)$, with $1 \le i \le n_{\rm r}$,
comprising each 1 (for $i=1$, i.e. at the origin)
or $6(i-1)$ sites, adding up to a total of
$n=7$, 19,~\ldots, 271 sites 
for $n_{\rm r} =2, 3,~\ldots, 10$.
The grid has six-fold symmetry around the axis orthogonal to the disc, 
and along each of the in-plane
symmetry axes, hard spheres on neighbouring sites overlap whenever
the aspect ratio $D/\sigma < 2(n_{\rm r}-1)$.
For small numbers of rings, it is clear that the resulting geometric object 
is  very different from a convex body, since the interaction-site platelet 
will have holes, or a highly corrugated surface, depending on 
the relative values of $D/\sigma$ and $n_{\rm r}$.
For any given $D/\sigma$, the platelet constructed from concentric
rings of spheres will tend to a cylinder of the same aspect ratio
only in the limit $n_{\rm r} \to \infty$.

The hexagonal grid shown in Fig.~\ref{fig:model} is identical
to that or Ref.~\cite{kutt:00}. Concentric hexagons replace the rings,
and the resulting total numbers of sites are identical to those obtained
with the circular pattern. The polygonal shape is reminiscent
of that of Gibbsite platelets used in recent experiments
at the Van't Hoff Laboratory in Utrecht~\cite{koij:98, koij:00}.

The interaction between two platelets is purely steric,
and consists of the sum of hard sphere interactions between all the pairs
of sites on the platelets, say $A$ and $B$:
\be\label{eq:pot}
v_{AB}=\sum_{i \in A} \sum _{j \in B} 
v(|{\mathbf r}_i-{\mathbf r}_j|)
\ee
where
${\mathbf r}_i$ and ${\mathbf r}_j$ are the positions
of sites $i$ and $j$, with $1 \le i,j\le n$ on platelets $A$ and $B$,
while $v(r)$ is the familiar hard sphere ``potential'':
\ba\label{eq:hs}
\begin{array}{cc}
v(r) = &  
\left\{
 \begin{array}{cc}
 \infty   & \quad {\rm if} \ r \le \sigma \cr
 0        & \quad {\rm if} \ r >  \sigma
 \end{array}
 \right. 
\end{array} 
\ea
This completely defines the interaction site model of platelets which 
we have investigated. The model
is obviously athermal, and its reduced thermodynamic properties
will only depend on the platelet density $\rho=N/V$.
The much-studied Laponite platelets have typical dimensions
$D\simeq 25$\,nm and $L$ (identified here
with the diameter $\sigma$ of the sphere) $\simeq 1$\,nm,
corresponding to an aspect ratio $D/\sigma=25$~\cite{lapo:90}.
The results which 
will be presented here are for aspect ratios $5 \le D/\sigma \le 50$,
as well as for infinitely thin platelets ($\sigma=0$).

\section{RISM and PRISM theories}

The pair structure of a fluid of identical particles, each carrying 
$n$ distinct interaction sites, is characterized
by a set of $n(n+1)/2$ site-site intermolecular
pair correlation functions $h_{ij}(r)=g_{ij}(r)-1$. These are
related to a set of $n(n+1)/2$ intermolecular direct correlation
functions $c_{ij}(r)$ by the RISM Ornstein-Zernike (OZ)
relations~\cite{chan:72}, which are most conveniently written
in matrix form and in Fourier space as:
\ba\label{eq:rism}
{\bf h}(k) = {\bf w}(k){\bf c}(k){\bf w}(k) + 
\rho{\bf w}(k){\bf c}(k){\bf h}(k)\,, 
\ea
where ${\mathbf h}\equiv [h_{ij}(k)]$, ${\mathbf c}\equiv[c_{ij}(k)]$, 
and ${\mathbf w}$
are $ n \times n$ symmetric matrices. 
The elements of ${\bf w}\equiv [w_{ij}(k)]$ 
are the Fourier transforms
of the intramolecular correlation functions. 
Provided the molecules are rigid,
we have explicitly:
\ba\label{eq:intra}
w_{ij}(k) = \frac{\sin(kL_{ij})}{kL_{ij}}\,,
\ea
where $L_{ij}$ is the bond length
between sites $i$ and $j$ on the same particle.
The RISM-OZ system~(\ref{eq:rism}) must be supplemented by
a set of closure relations relating the $h_{ij}$ and
$c_{ij}$. Since only hard-core interactions are present,
the simplest choice is the Percus-Yevick (PY) closure, which
combines the exact core condition
\be\label{eq:core}
h_{ij}(r) = -1\,, \qquad r \le  \sigma
\ee
with the approximation
\ba\label{eq:py}
c_{ij}(r)  =  0\,, \qquad r >  \sigma \,.
\ea

The number of independent $h_{ij}$
and $c_{ij}$ may be drastically reduced by obvious symmetry
considerations. In the circular platelet model of Fig.~\ref{fig:model}
all sites on a given ring are equivalent, so that the total number
of independent correlation functions
is reduced from $n(n+1)/2$ to $n_{\rm r}(n_{\rm r}+1)/2$.
Following Raineri and Stell~\cite{rain:01},
we introduce the symmetry-reduced 
matrix ${\bf W}(k)$ of 
intramolecur correlation functions 
with elements:
\ba\label{eq:redw}
{\bf W}_{IJ}(k)& = & \frac{1}{n_J} \sum_{j \in J} w_{ij}(k) \nonumber\\
               & = & \frac{1}{n_I} \sum_{i \in I} w_{ij}(k) ={\bf W}_{JI}(k)
\ea
where the indices $I$ and $J$ run over the number of inequivalent 
classes of interaction sites and $n_I$ or $n_J$ 
are the numbers of equivalent sites of classes $I$ or $J$
in the molecule.
We also define the following direct correlation function 
matrix $\bf C$ with elements:
\be\label{eq:cij}
C_{IJ}(k)=n_I\cdot c_{ij}(k)\cdot n_J
\ee
where $c_{ij}(k)$ is any of the identical direct correlation function
of the original $n \times n$ matrix, with $i \in I$ and $j \in J$.
The reduced RISM-OZ equation now reads in $n_{\rm r} \times n_{\rm r}$
matrix form:
\ba\label{eq:redrism}
{\bf h}(k) = {\bf W}(k){\bf C}(k) {\bf W}(k) +
\rho{\bf W}(k) {\bf C}(k) {\bf h}(k)\,,
\ea
which is once more supplemented by the PY closure
relations~(\ref{eq:core}) and~(\ref{eq:py}).
In the case of the hexagonal grid in Fig.~\ref{fig:model}, 
the reduction procedure due to symmetry is slightly
more elaborate. Consequently, we made the simple (but slightly inaccurate)
assumption that all sites on a given hexagonal shell (also referred
to as ``ring'')  are equivalent, which is strictly true
only for the $n_{\rm r}=2$ shell.

Despite its diagrammatic inconsistency~\cite{chan:82},
RISM has proved to be a successful theory of the pair structure
of many simple molecular fluids~\cite{mons:90}.
In the case of macromolecular and colloidal systems,
with very large numbers of interaction sites, the number of coupled
RISM equations become intractable, and a considerable simplification
follows from the assumption that all interaction sites are equivalent.
This leads to the PRISM theory, first applied
by Schweizer and Curro to long flexible polymers~\cite{schw:87}.
PRISM theory neglects end effects in that case. In the case of 
rigid lamellar particles of finite diameter, the PRISM
assumption is clearly less justified, particularly
so as the number $n_{\rm r}$ of concentric rings
of sites increases. The set of correlation
functions in Eq.~(\ref{eq:rism}) is replaced by a single,
platelet-averaged function $h$, defined by:
\ba\label{eq:hprism}
h(k)=\frac{1}{n^2}\sum_{i,j} h_{ij}(k)
\ea
and similarily for $c(k)$. The platelet-averaged intramolecular 
correlation function is defined by:
\ba\label{eq:wprism}
w(k)=\frac{1}{n}\sum_{i,j}w_{ij}(k)\,,
\ea
and the resulting single PRISM-OZ relation reads
\ba\label{eq:prism}
h(k)=w(k)c(k) [w(k)+\rho n h(k)]\,,
\ea
where $\rho n$ is the number density of interaction sites.
Note that $w(k)$ is proportional
to the intramolecular form factor $F(k)$
of a single platelet:
\ba\label{eq:fk}
F(k) & = & 
\frac{1}{n} w(k) = 
\frac{1}{n^2}\sum\limits_{i,j} w_{ij}(k)  \nonumber\\
& = & \frac{1}{n} + 
\frac{1}{n^2}\sum\limits_{i\ne j} w_{ij}(k)  \,.
\ea
which characterizes the geometry of the distribution of sites, and hence the
geometric shape of the model platelets. While the form factor 
accounts for 
the interference of radiation scattered from
different parts of the same particle in an X-ray or
neutron diffraction experiment, the local order in the fluid
or suspension is characterized by the set of pair distribution
functions $g_{ij}(r)$, or, approximately, by the pre-averaged
single combination~(\ref{eq:prism}). An important particular case
is the centre-to-centre pair distribution function $g_{ \rm cc}(r)$.
The experimentally accessible total structure 
factor is:
\ba\label{eq:sk}
S(k)=F(k)+
\frac{1}{n^2}\sum\limits_{i,j} S_{ij}(k)
\equiv F(k)+S_{\rm inter}(k)\,,
\ea
where 
the site-site partial structure factors $S_{ij}(k)$ 
are the Fourier transforms
of the $g_{ij}(r)$. 
The structure factor
provides a direct link with
thermodynamics, via the compressibility
equation~\cite{hans:86}:
\ba\label{eq:comp}
\lim\limits_{k \to 0} S(k)=\rho k_{\rm B} T \chi_T\,,
\ea
where $\chi_T$ is the isothermal compressibility.
The osmotic pressure then follows from:
\ba\label{eq:press}
\beta P \equiv \frac{P}{k_{\rm B}T} = 
\int_0^\rho d\rho^\prime \, [S(k=0)]^{-1}\,.
\ea

Returning to PRISM theory, the OZ relation~(\ref{eq:prism})
must be supplemented by a closure relation for the 
pre-averaged functions $h$ or $c$. In this
work we use throughout the PY closure~(\ref{eq:core}) and~(\ref{eq:py}),
both within the RISM and PRISM representations. 
PRISM theory has so far been mostly applied to polymer solutions and melts,
whence the name~\cite{schw:97}. For flexible
polymers the form factor $F(k)$ follows from
a statistical average over single polymer conformations,
while in the case of rigid rods or platelets considered in 
the present work, $F(k)$, or equivalently $w(k)$,
are trivially known functions of the wave number $k$.
The case of rigid rods
was investigated within the PRISM
formalism both
as the limit of polymers with a persistence 
length exceeding the overall polymer length~\cite{pick:00},
and by using
the correct form factor (see~\cite{harn:00}
and references therein).


\begin{figure}
\begin{center}
\includegraphics[width=6.5cm,angle=-90]{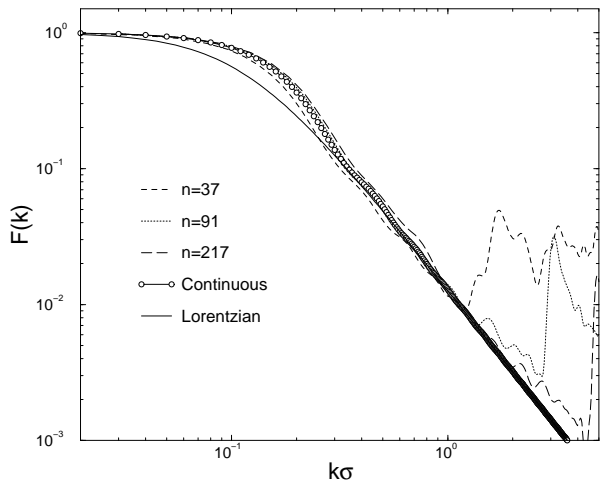}
\caption{
Form factor of model platelets, Eq.~(\ref{eq:fk}),
with $D/\sigma=25$ and
different numbers $n$ of sites,
compared with the continuous distribution
limit~(\ref{eq:fkcont}) and the Lorentzian
approximant~(\ref{eq:wlorentz}).
Symbols are defined in the figure.
}\label{fig:fk}
\end{center}
\end{figure}

\section{Pair structure}

We have solved the RISM and PRISM equations for various interaction site
distributions,
with number of sites $n$ ranging from $n=37$ ($n_{\rm r}=4$)
to $n=271$ ($n_{\rm r}=10$), for aspect ratios $D/\sigma$
ranging from 5 to 25 and in the limit $\sigma=0$.
Test calculations with $D/\sigma=50$ and $n_{\rm r}=18$ have also
been carried out.

The advantage of using
symmetry-reduced RISM and PRISM theories is that
$n$ may be increased
within a reasonable computational effort,
in order to mimic, as accurately as possible, 
a uniform distribution over the platelet area,
and hence a smooth surface of a thin cylinder.
In fact, the limit of 
a continuous distribution may be achieved straightforwardly 
in the PRISM formalism. 

\begin{figure}
\begin{center}
\includegraphics[width=6.5cm,angle=-90]{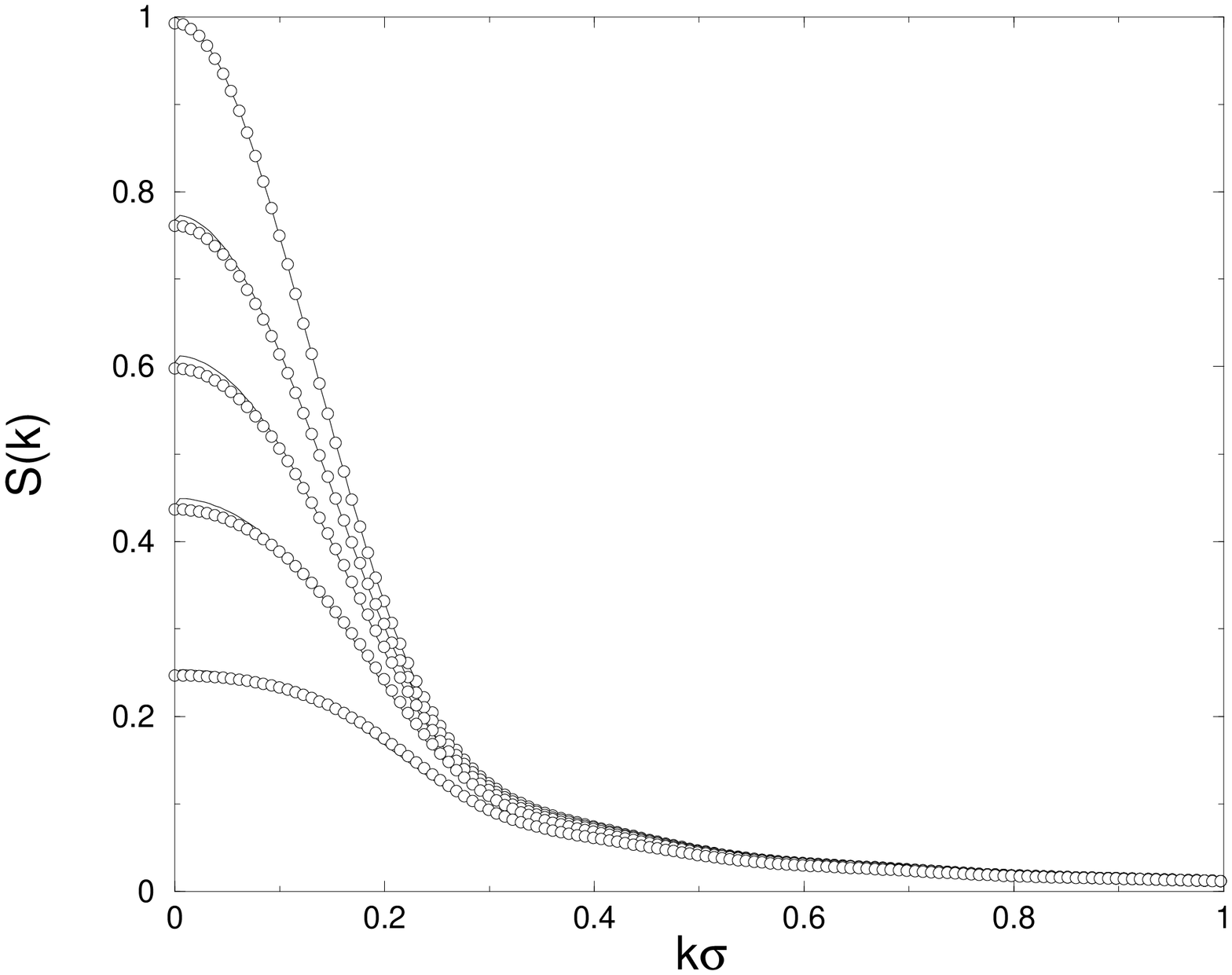}
\includegraphics[width=6.5cm,angle=-90]{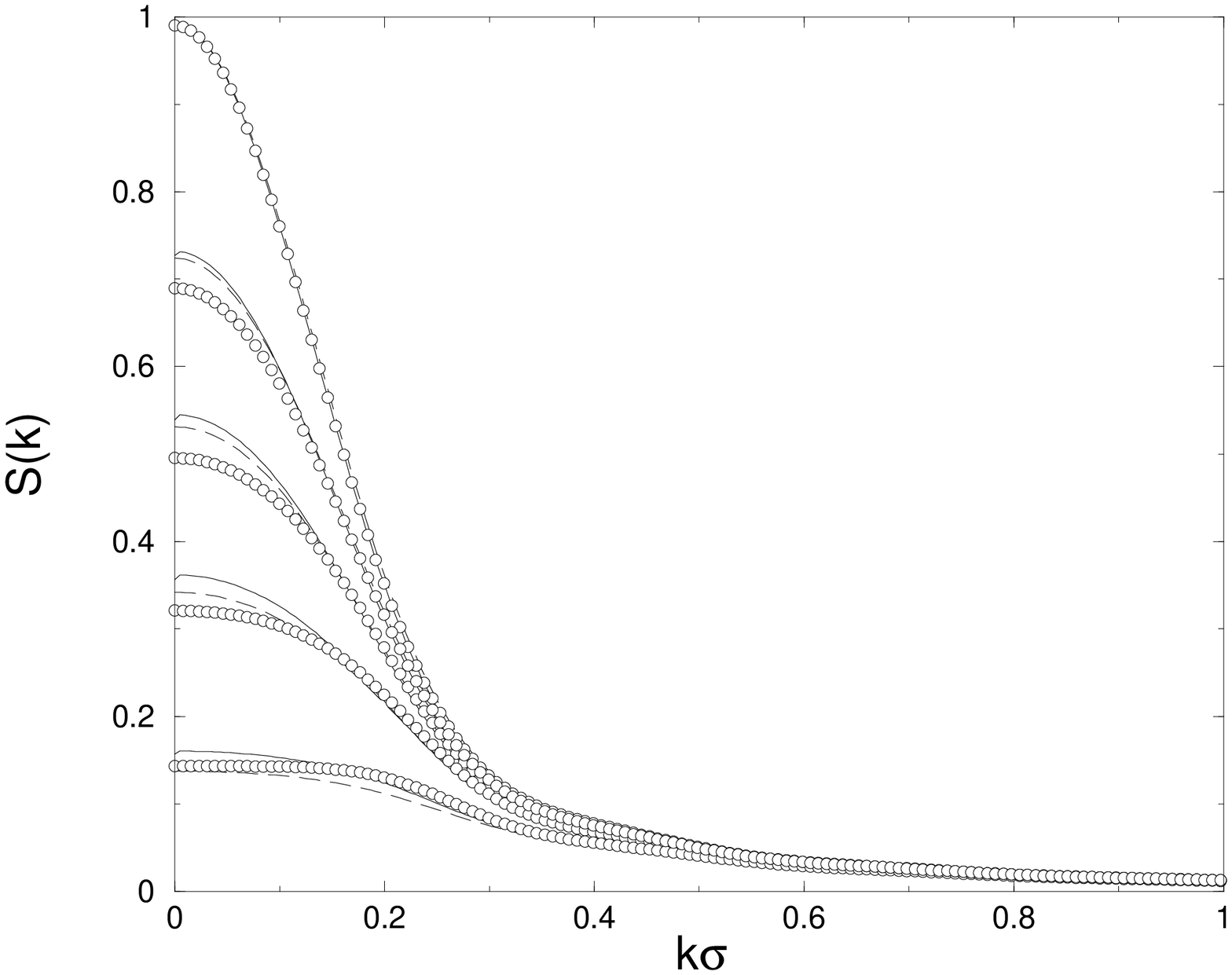}
\caption{Total structure factor for model platelets
with $D/\sigma=25$, and 91 (top) or 217 (bottom) sites.
Circles: RISM;  lines: PRISM results 
with the discrete [Eq.~(\ref{eq:fk}), full lines],
or the continuous [Eq.~(\ref{eq:fkcont}), dashed lines in the bottom panel]
form factors.
Reduced densities are $\rho^*=\rho D^3=0.01$, 0.40, 0.80, 1.40, 2.80
(from top to bottom).
}\label{fig:sk217}
\end{center}
\end{figure}

The importance of ``holes'' 
between sites on the grid 
may be seen in Fig.~\ref{fig:fk},
where the form factors $F(k)$, defined in Eq.~(\ref{eq:fk}),
are plotted
for several values 
of the number of sites, and compared  to the 
limit of a continuos distribution
of sites, when 
the sum in Eq.~(\ref{eq:fk}) goes  over 
to an integral on the surface of the platelet,
with the result~\cite{guin:55}:
\ba\label{eq:fkcont}
F(k)=\frac{2}{(kR)^2} \left[ 1-\frac{J_1(2kR)}{kR}\right]\,,
\ea
where $J_1$ denotes the cylindrical Bessel function
of order one.
The discrete nature of the site distribution
is reflected in the irregular  oscillations 
visible for $k\sigma>1$ in Fig.~\ref{fig:fk}. 
In this regime,
the correct $k^{-2}$ power law decay at large $k$
is progressively recovered by increasing the number $n$ of sites
for a given diameter $D$.
A model with $n_{\rm r}=9$ ($n=217$
sites), where the distance between
nearest-neighbour sites is $1.5\sigma$, 
reproduces the continuous form factor~(\ref{eq:fkcont})
with reasonable accuracy.
Figure~\ref{fig:fk} also displays
a simple Lorentzian approximation of the form factor,
suitable for analytical
calculation~\cite{harn:01,pick:99}: 
\ba\label{eq:wlorentz}
F(k) = \frac{2}{2+(kR)^2} \,.
\ea

\begin{figure}
\begin{center}
\includegraphics[width=6.5cm,angle=-90]{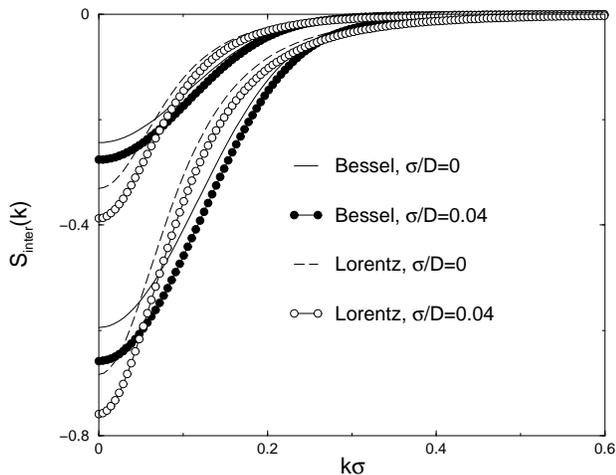}
\caption{Intermolecular structure factor $S_{\rm inter}(k)$,
for model platelets 
with $D/\sigma=25$ or $\infty$, calculated
within PRISM, with the
Lorentzian form factor [Eq.~(\ref{eq:wlorentz})], and the
Bessel form factor [Eq.~(\ref{eq:fkcont})].
Reduced densities are $\rho^*=\rho D^3=0.40$ (upper curves)
and 1.40 (lower curves).
}\label{fig:skbessel_inter}
\end{center}
\end{figure}

\begin{figure}
\begin{center}
\includegraphics[width=6.5cm,angle=-90]{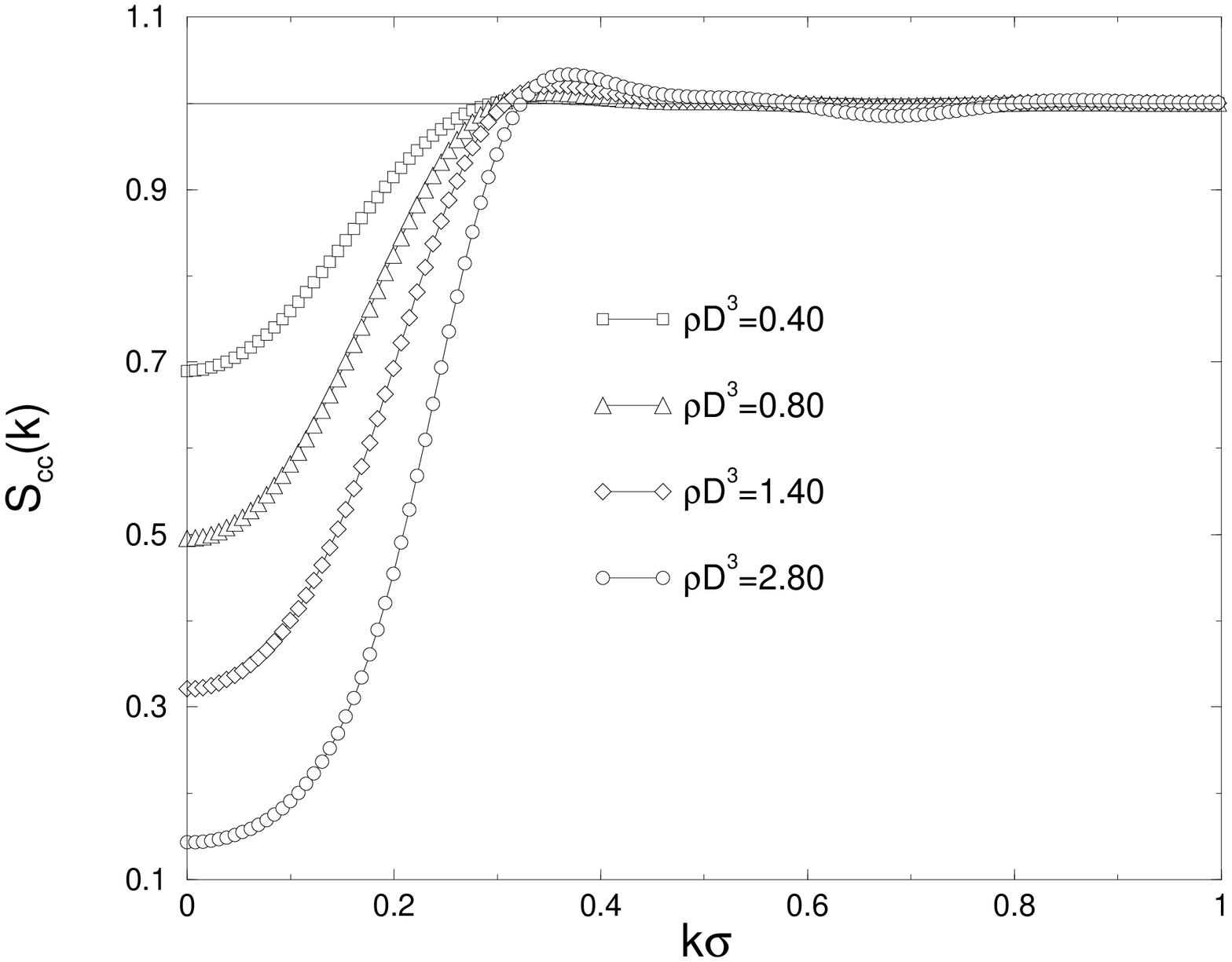}
\includegraphics[width=6.5cm,angle=-90]{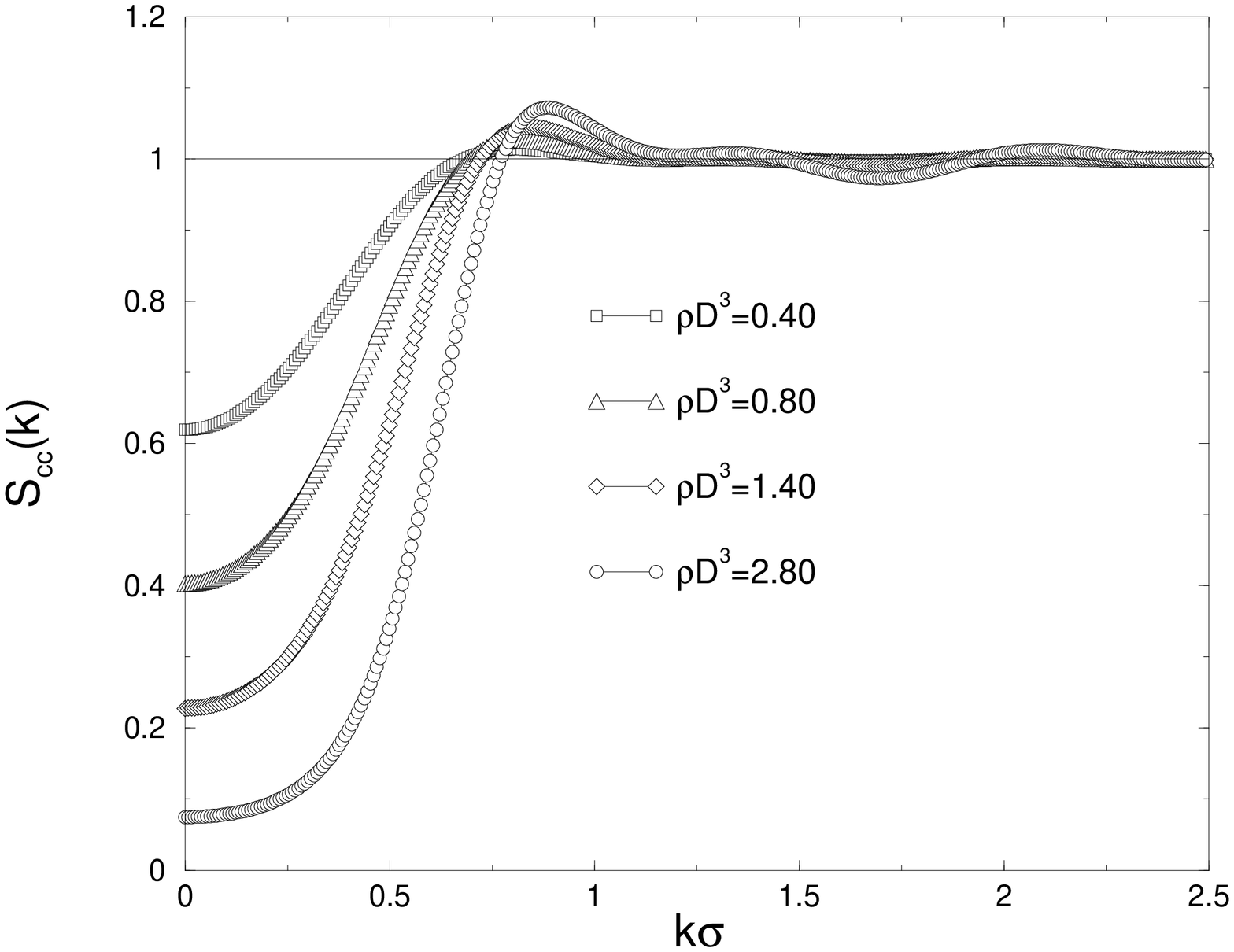}
\caption{RISM centre-to-centre structure factors $S_{\rm cc}(k)$,
for platelet models
with $D/\sigma=25$ (top) and 10 (bottom), for
densities specified by the symbols.
}\label{fig:sk_cc}
\end{center}
\end{figure}

Results for the overall structure factor~(\ref{eq:sk}), and
for the centre-to-centre structure factor obtained by solving the 
RISM and PRISM equations for the circular pattern
of sites shown in Fig.~\ref{fig:model} are presented 
in Figures~\ref{fig:sk217}-\ref{fig:sk_cc}.
For the hexagonal pattern, 
the full and approximate symmetry-reduced RISM calculations 
yield practically
indistinguishable results (not shown), and the
structure factors 
are very similar to those obtained 
with the circular pattern, for the same number of sites $n$, 
aspect ratio $D/\sigma$, and reduced density.
Figure~\ref{fig:sk217} compares the RISM and PRISM results for the total
structure factor at five different densities and for an aspect ratio
$D/\sigma=25$.
The same discrete form factor defined by Eqs.~(\ref{eq:intra})
and~(\ref{eq:fk}) was used in the two integral equations for the models
with $n=91$ and 217 sites.
In the $n=91$ case, RISM and PRISM results are virtually indistinguishable,
and this agreement is even better for smaller $n$.
The agreement deteriorates somewhat as $n$ increases, 
as illustrated by the case with $n=217$,
showing that 
the PRISM pre-averaging assumption gradually
breaks down when the number of sites increases.
As visible from Fig.~\ref{fig:sk217},
the PRISM structure factors are slightly modified by going 
to the continuum limit, 
characterized by the form factor~(\ref{eq:fkcont}), and 
move closer to the RISM data, pointing to some degree of 
cancellation of errors.

A comparison of the intermolecular part 
of the structure factors 
$S_{\rm inter}(k)$ [see Eq.~(\ref{eq:sk})],
calculated within PRISM in the limit of a continuous
distribution of sites ($n \to \infty$) using the exact 
[Eq.~(\ref{eq:fkcont})], and the approximate [Eq.~(\ref{eq:wlorentz})],
form factors is made in Fig.~\ref{fig:skbessel_inter}
for size ratios $\sigma/D=0$ and 0.04; the results
are seen to be very sensitive to the choice of form factor,
particularly at small wavenumbers $k$, which 
will have implications for the equation of state 
calculated via Eqs.~(\ref{eq:comp}) and~(\ref{eq:press}).

Note that the only justification for the use of the approximate 
Lorentzian form factor
is that it allows an analytic solution of the PRISM
equation in the limit of infinitely thin ($\sigma/D \to 0$)
hard 
as well as charged platelets~\cite{harn:01}.
Indeed, in that limit
the PRISM equation~(\ref{eq:prism}),
supplemented by the closure relation:
\be\label{eq:clorentz}
c({\mathbf r}) = c_0 \delta({\mathbf r})
\ee
can then be solved analytically~\cite{pick:99}; the parameter $c_0$
is determined by enforcing the exact core condition $h({\mathbf r}=0)=-1$.
The resulting structure factor reads:
\ba\label{eq:sklorentz}
S(k)  & = &    w(k) + \rho h(k) \nonumber \\[4pt] 
& = &   2\left\{2+2\rho A(R)
+ (kR)^2 \right\}^{-1} \\[4pt]
A(R) & = & 
8\pi \left(\frac{R}{\sqrt{2}}\right)^3 \
+  16\pi^2\rho \left(\frac{R}{\sqrt{2}}\right)^6  \nonumber \,.
\ea
The result~(\ref{eq:sklorentz})
is, not surprisingly, identical to that obtained from the PRISM solution
of the thread model, using the same closure~(\ref{eq:clorentz})
and the same approximant~(\ref{eq:wlorentz})
to the form factor, with the radius $R$ of the platelet
replaced by the radius of gyration $R_{\rm g}$
of the polymer~\cite{pick:99}. This 
coincidence illustrates the importance of using the correct 
form factor of the objects under consideration. 
The
form factor $F(k)$ for the platelets is known 
exactly [cfr Eq.~(\ref{eq:fkcont})],
but only approximately in the case of polymers, where
it is often taken to be of the Debye form, valid for ideal (gaussian)
polymer coils~\cite{barr:03}.

\begin{figure}
\begin{center}
\includegraphics[width=6.5cm,angle=-90]{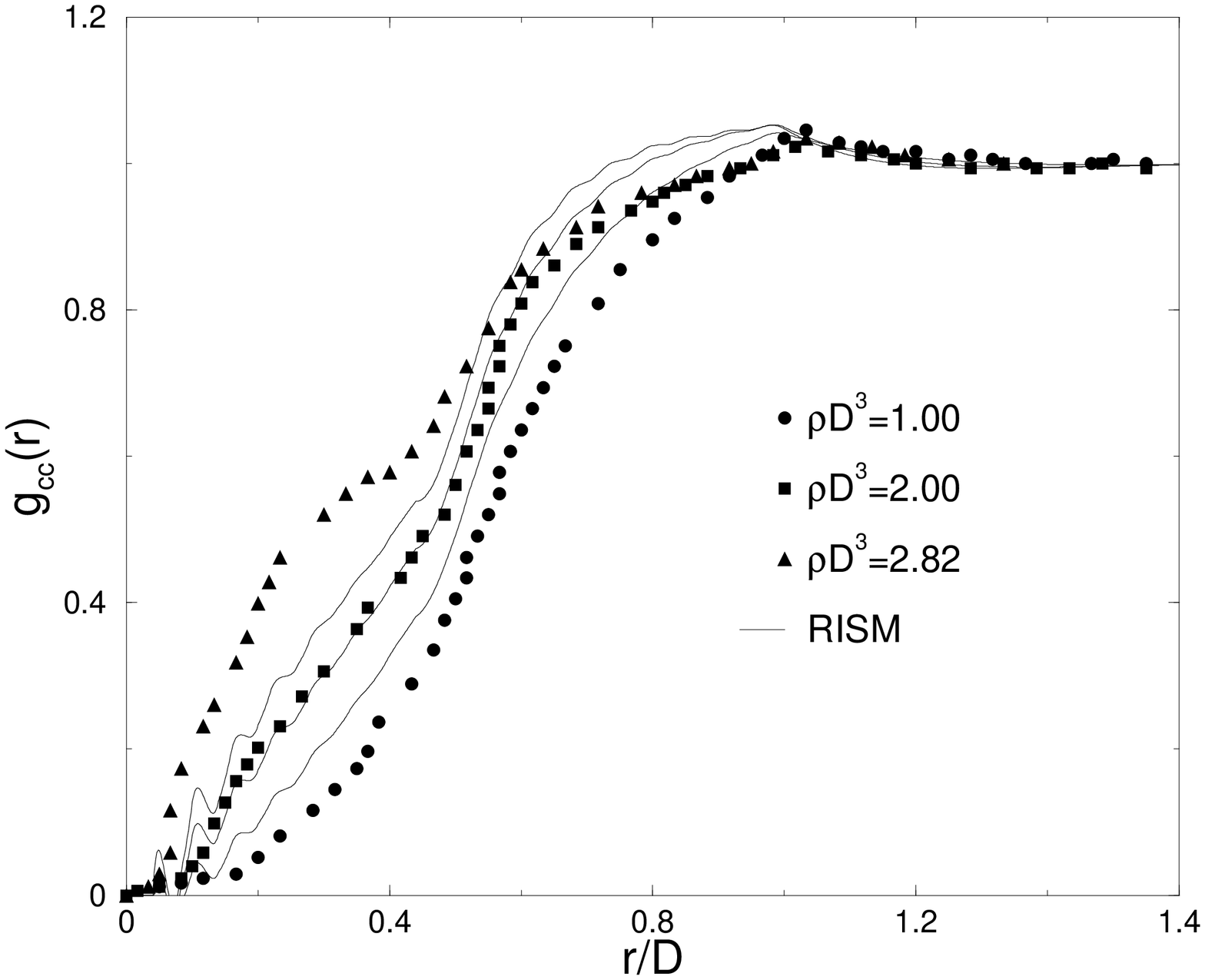}
\includegraphics[width=6.5cm,angle=-90]{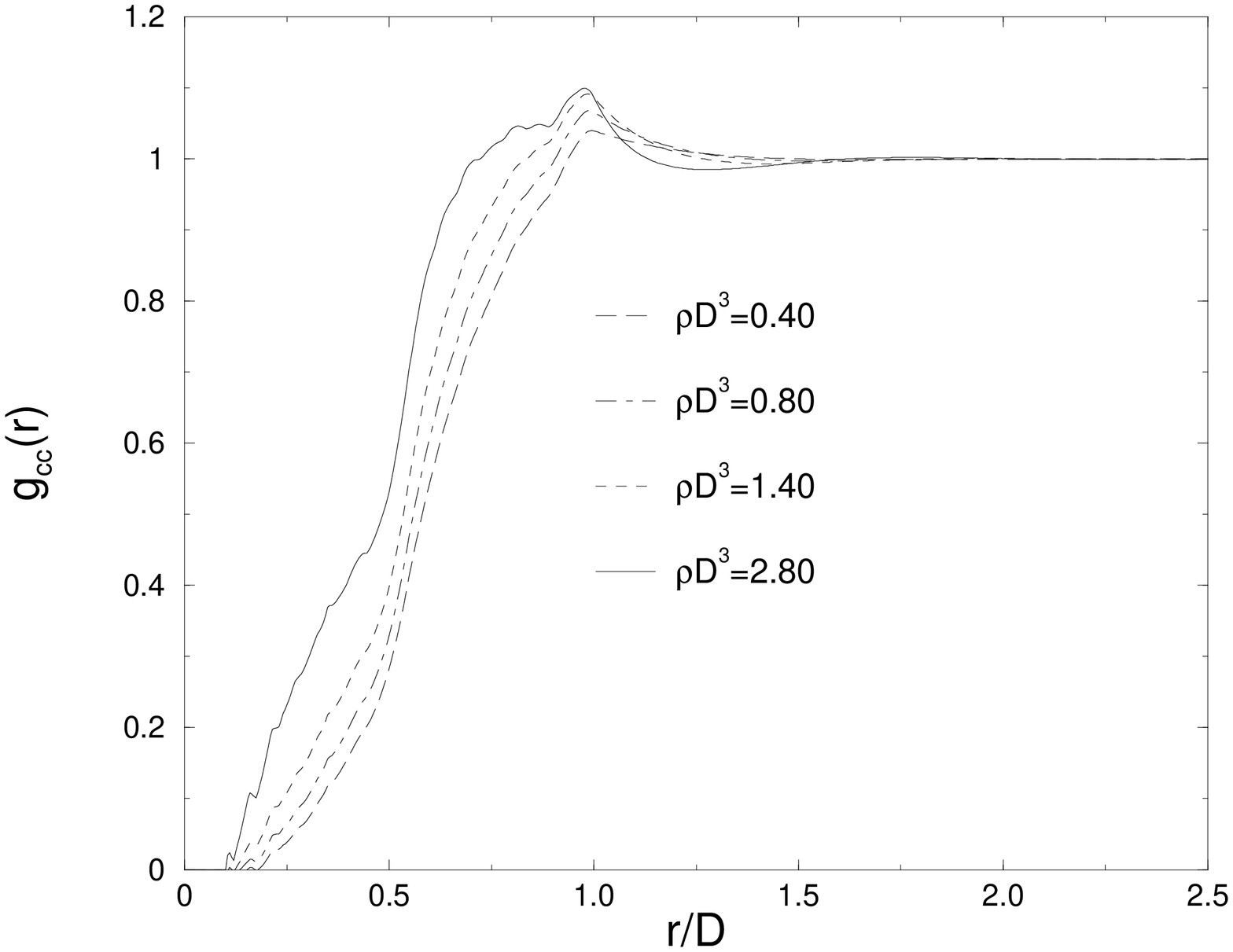}
\caption{Top: RISM centre-to-centre radial distribution function for
platelets with $D/\sigma=25$
(lines, in order of increasing densities, from left to right). 
Symbols are the corresponding
simulation data for infinitely thin 
hard discs~\protect\cite{eppe:84}.
Bottom: RISM radial distribution functions
for platelets with $D/\sigma=10$.
}\label{fig:rdf}
\end{center}
\end{figure}

RISM results for the centre-to-centre structure factor
$S_{\rm cc}(k)$, calculated for the $n=217$ sites model, are
shown in Fig.~\ref{fig:sk_cc} for two size ratios, $D/\sigma=10$ and 25, 
and four densities.
As expected, the $S_{\rm cc}(k)$ for the smaller size ratio show 
more structure because the corresponding effective 
packing fractions  are larger.
The corresponding centre-to-centre pair distribution function, 
$g_{\rm cc}(r)$, for $D/\sigma=25$, is shown in Fig.~\ref{fig:rdf},
and compared to Eppenga and Frenkel's Monte Carlo (MC) data for infinitely
thin platelets~\cite{eppe:84}.
The oscillations in the RISM results at short distances
reflect the surface ``roughness'' 
due to the discrete site distribution, and are of course 
absent in the MC data.
Bearing in mind the differences between the two models, 
the agreement between theory and simulation is reasonably good, 
and all curves exhibit the expected cusp at $r \simeq D$.
At the highest density the MC data exhibit a steep 
rise followed by a striking inflection point around $r \simeq 0.4D$,
which is only faintly reproduced by RISM; for $r \lesssim 0.4D$,
the MC $g_{\rm cc}(r)$ lies well above the RISM prediction, pointing
to some degree of local orientational ordering, which may be
regarded as a precursor to the isotropic-nematic transition
taking place at $\rho D^3 \simeq 4$~\cite{eppe:84}.
Pair distribution functions $g_{\rm cc}(r)$ for the size ratio
$D/\sigma=10$ are reported in the bottom panel of Fig.~\ref{fig:rdf}
for four densities. The qualitative behaviour is similar
to that predicted by RISM for the larger size ratio; a similar 
cusp appears at $r \simeq D$.

\section{The equation of state}

From a knowledge of the site-site correlation functions, the only  direct
route to the osmotic equation of state is via Eqs.~(\ref{eq:comp}) 
and~({\ref{eq:press}). Results from RISM and PRISM calculations for
circular site patterns with $ 37 \le n \le 271$  sites,
and an aspect ratio $D/\sigma=25$ are shown in Fig.~\ref{fig:press},
and compared to the MC data for infinitely thin
platelets~\cite{eppe:84}.
For any given reduced density $\rho^*=\rho D^3$, the pressure
increases with the number $n$ of sites. This is easily
understood, because for lower values of $n$, the distance
between neighbouring sites exceeds  $2\sigma$, so that
two platelets can, in fact, partially interpenetrate for
favourable relative orientations. But
for $n=217$, the distance between neighbouring spheres ($\simeq 1.5\sigma$) 
is small enough to avoid interpenetration, and the value
of the pressure saturates, as was checked by RISM test calculations
with ten concentric rings, corresponding to  $n=271$ sites.
These saturated values of the pressure lie about 10\% below
the MC data in the intermediate density range. Interestingly, 
the MC data show very little curvature
of the equation of state for $\rho^* \gtrsim 1.5$,
and fall below the predictions of RISM and of the fifth order
virial expansion at higher densities. This ``softening'' of
the equation of state may be indicative of local orientational
ordering prior to the isotropic-nematic transition which occurs 
at $\rho^* \simeq 4$. Such local ordering, already apparent
 in the pair distribution function shown in Fig.~\ref{fig:rdf}, 
 cannot be accounted for by RISM theory,
 which assumes full isotropy of the local correlations.
 The PRISM results, obtained with the discrete 
 form factor~(\ref{eq:wprism}) appropriate
 for a given number of sites, are in excellent agreement with the RISM
 pressures up to $n=91$. Deviations
 between PRISM and RISM
 results become apparent for $n=127$ (not shown) 
 and increasingly so for $n=217$.
 However, if the continuum limit ($n \to \infty$)
 is taken within PRISM, the form factor
 goes over to that given in Eq.~(\ref{eq:fkcont}),
 and the resulting equation of state agrees then closely with the RISM
 results for $n=217$, which are close to
 saturation, i.e. to the expected continuum limit
 within RISM.

\begin{figure}
\begin{center}
\includegraphics[width=6.5cm,angle=-90]{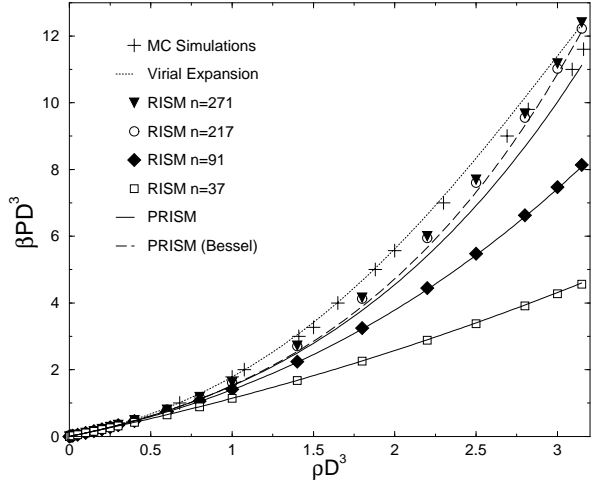}
\caption{Equation of state of hard 
platelets with $D/\sigma=25$ 
in the isotropic fluid range, as obtained through
RISM (symbols) and PRISM (full lines)
calculations with $n=37$, 91, and 217 sites and the {\it same} 
form factor of Eq.~(\ref{eq:fk}).
RISM test calculations with $n=271$, 
and PRISM results with the continuous 
Bessel form factor, Eq.~(\ref{eq:fkcont}),
are also reported.
Monte Carlo
data and the 5th order virial
expansion for infinitely thin hard platelets
are taken form Ref.~\protect\cite{eppe:84}. 
}\label{fig:press}
\end{center}
\end{figure}

\begin{figure}
\begin{center}
\includegraphics[width=6.8cm,angle=-90]{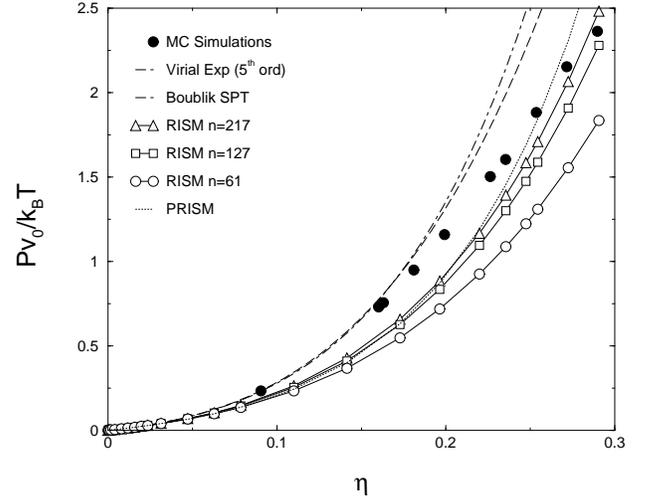}
\caption{RISM and PRISM (with 
Bessel form factor) predictions for the equation of state of model
platelets with $D/\sigma=10$.
Results are compared with Monte Carlo simulations of cut hard spheres
of Ref.~\protect\cite{veer:92}.
}\label{fig:pressother}
\end{center}
\end{figure}

 Similar RISM and PRISM results for the equation
 of state of discs with an aspect ratio $D/\sigma=10$
 are compared in Fig.~\ref{fig:pressother}
 to MC data for ``cut hard spheres'' with the same aspect 
 ratio~\cite{veer:92}. The trends are qualitatively the same
 as for the previous case with $D/\sigma=25$. The
 ``softening'' of the ``exact'' equation of state (as
 obtained by MC simulations) is again very striking
 at packing fractions $\eta \gtrsim 0.2$, where the MC data
 are bracketed by the predictions of the 5th order
 virial expansion (which lie above)  and the
 RISM predictions. Close to the expected
 isotropic-nematic transition at $\eta \simeq 0.3$~\cite{veer:92}
 the ``saturated'' RISM pressure 
 (estimated from calculation for the $n=217$ sites model)
 nearly coincides with the MC value. The continuum PRISM
 equation of state is close to the ``saturated'' RISM 
 result as in the case with $D/\sigma=25$. 

 The PRISM pressures 
 for $D/\sigma=5$, 10, 15, 20, 25, 50 and in the limit $\sigma \to 0$ 
 are collected
 in Fig.~\ref{fig:pvsb2}, where the platelet density
 and the ratio $P/k_{\rm B}T$ are both reduced by the second 
 virial coefficient in the isotropic phase~\cite{onsa:49}:
 \be\label{eq:b2}
 B_2= \frac{\pi^2}{16}D^3 + \frac{\pi}{8}(3+\pi)D^2\sigma +
 \frac{\pi}{4}D\sigma^2 \,.
 \ee
When $P^*=\beta PB_2$ is plotted versus $\rho^*=\rho B_2$,
the data nearly collapse on a single ``master curve'',
which is reasonably well fitted by a cubic polynomial.
Only the results for $D/\sigma=5$ lie significantly
above the master curve for $\rho^*> 1$. 

\begin{figure}
\begin{center}
\includegraphics[width=6.8cm,angle=-90]{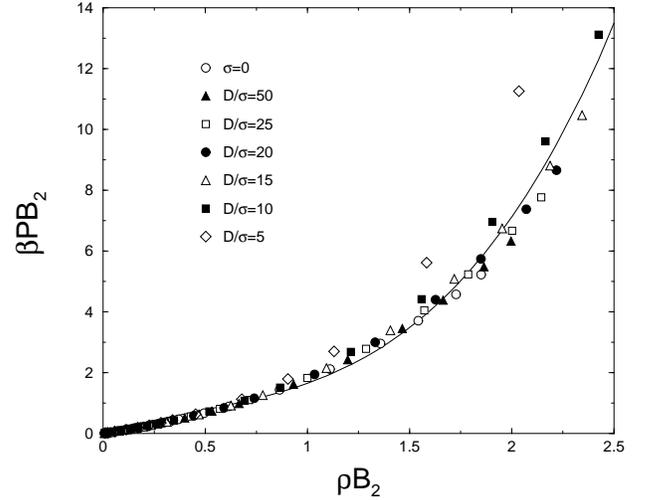}
\caption{
Equation of state of model platelets in reduced units of
pressure and density, $P^*=\beta PB_2$, $\rho^*=\rho B_2$
(where $B_2$ is the second virial coefficient).
Full line is a cubic fit of all data in the range 
$10 \le D/\sigma < \infty$.
}\label{fig:pvsb2}
\end{center}
\end{figure}

Equations of state obtained with the circular
and the hexagonal grids of Fig.~\ref{fig:model} are
reported in Fig.~\ref{fig:pvvseta_hex}
for $D/\sigma=25$, and $n=37$, 91, and 217 sites.
As visible,
the pressures of both models agree closely, 
if the densities $\rho$ 
are properly reduced by the volume of a circular or
hexagonal platelet in the continuum ($n \to \infty$)
limit. 

We finally return to the prediction
of the continuum PRISM theory, using the PY
closure for finite thickness
$D/\sigma=25$ and 50, and the closure~(\ref{eq:clorentz})
for infinitely thin ($\sigma \to 0$) platelets. 
PRISM results, obtained with the exact~(\ref{eq:fkcont}) 
and the approximate~(\ref{eq:wlorentz})
form factors
are reported in Fig.~\ref{fig:lrntz}, and compared
to the RISM predictions for $D/\sigma=25$ and 50 
(with $n_{\rm r}=9$ and 18, respectively).
Two main conclusions which may be drawn from the comparison
are the following. 
First,
the pressures are higher
when $\sigma=0.04D$, compared to $\sigma=0.02D$ (and $\sigma=0$
within PRISM), as expected for
obvious excluded volume reasons, but the effect
appears to be exagerated by PRISM theory. 
Finally,
as already evident from the comparison
of the PRISM structure factors shown in Fig.~\ref{fig:skbessel_inter},
the form factor has a considerable
influence on the PRISM results with the Lorentzian
approximant to the exact form factor always
leading to an overestimation of the pressure.
In view of the present discussion, the good agreement
between the predictions of
the continuum PRISM theory, obtained with the Lorentzian
form factor~(\ref{eq:wlorentz}), and the ``exact'' MC data for 
infinitely thin platelets~\cite{eppe:84} noted
in Ref.~\cite{harn:01}, must
be considered as partly accidental, due
to a cancellation of errors arising from the
combination of an inadequate form factor 
and the assumption of equivalence of all sites.

\begin{figure}
\begin{center}
\includegraphics[width=6.5cm,angle=-90]{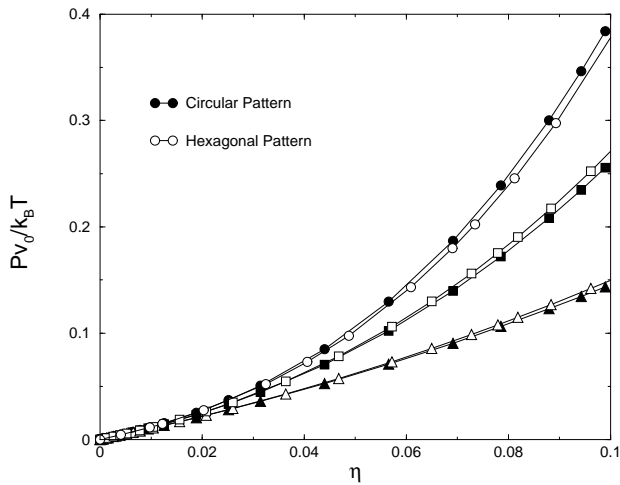}
\caption{RISM equation of state
for model platelets with different geometries,
$D/\sigma=25$ and
$n=217$, 91, or 37 sites (from top to bottom).
Full symbols: circular geometry;
open symbols: hexagonal geometry (see Fig.~\ref{fig:model}). 
$\eta=\rho v_0$ is the dimensionless packing fraction
and $v_0$ is the particle volume.
}\label{fig:pvvseta_hex}
\end{center}
\end{figure}

\section{Conclusions}

We have presented results of what we believe to be the first
systematic application of RISM and PRISM theories
to rigid, lamellar colloidal particles.
While RISM has previously been almost exclusively applied to rigid
molecules~\cite{mons:90}, and PRISM, as indicated by the first letter
of the acronym, has been mostly applied to polymers~\cite{schw:97}, our
investigation presents a systematic comparison of
the predictions of the two (related)
theories as applied to colloidal platelets.
In this paper we have restricted our attention to systems with 
excluded volume (hard core) interactions only between the 
sites distributed in a circular or hexagonal pattern. 
For low numbers $n$ of sites, the model platelets have many 
``holes'' or exhibit highly corrugated surfaces, but for sufficiently 
large $n$, the spheres associated with neighbouring sites touch 
or overlap, and in the continuum limit $n \to \infty$, which 
can easily be dealt with within the PRISM theory, the topology of 
the multi-site plates goes over to smooth cylindrical or hexagonal bodies. 

\begin{figure}
\begin{center}
\includegraphics[width=6.5cm,angle=-90]{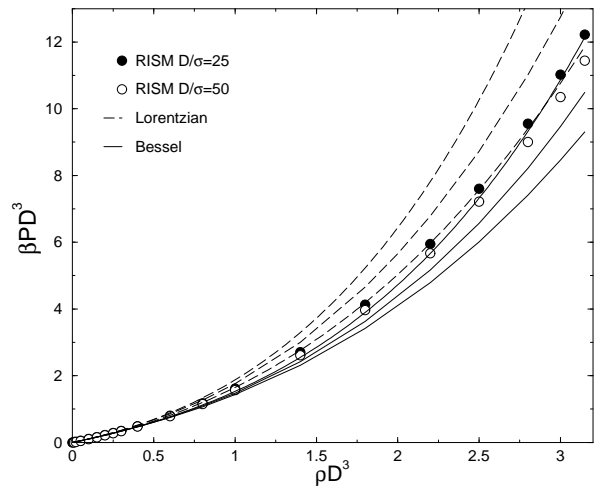}
\caption{Equation of state of thin model platelets. 
RISM (full circles: $D/\sigma=25$, open circles: $D/\sigma=50$);
PRISM with the Lorentzian form factor, Eq.~(\ref{eq:wlorentz})
(dashed lines, $D/\sigma=25$, 50 and $\infty$, from top to bottom);
PRISM with the exact Bessel form factor, Eq.~(\ref{eq:fkcont})
(full lines, $D/\sigma=25$, 50 and $\infty$, from top to bottom).
}\label{fig:lrntz}
\end{center}
\end{figure}

The main findings and conclusions of the present work may be 
summarized as follows: \par
(a) RISM theory predicts reasonable pair structure, as embodied 
in the centre-to-centre pair distribution function (cfr Fig.~\ref{fig:rdf})
or the total structure factor
(cfr Fig.~\ref{fig:sk217}), compared to available simulation data.
The fully isotropic form factor used in all our calculations, 
and the ensuing complete decorrelation between plate orientations and 
the site-site correlation functions, do not allow a proper description 
of the local orientational ordering observed in the MC simulations. \par
(b) The RISM equation of state converges rather slowly to the 
hard platelet limit, as the number of interaction sites increases, 
but ``saturation'' is reached when the spheres associated with 
neighbouring sites are close enough
to prevent the interpenetration of platelets. 
The limiting equation of state is strongly 
convex and shows no sign of the ``softening'' observed in MC simulations 
well before the isotropic-to-nematic transition is reached. 
The RISM pressures lie systematically below the predictions of a 
five-term virial series, and of the closely related Boublik equation 
of state.  \par
(c) The properly scaled RISM equation of state data 
over a wide range of aspect ratios $D/\sigma$ 
collapse onto a 
single ``master curve''. \par
(d) The results of the much simpler PRISM theory, which assumes 
all interaction sites to be equivalent, agree unexpectedly well 
with their RISM counterparts, when the appropriate form factors are used. 
In other words, pre-averaging effects do not seem to play an important role. 
The continuum ($n \to \infty$) limit of PRISM
theory performs remarkably well, considering its simplicity and 
modest computational demands, provided the corresponding exact 
form factor~(\ref{eq:fkcont}) is used. The pressures obtained when the 
Lorentzian approximant (17) to the form factor is used
in the PRISM calculations, are systematically too high.
The substantial differences observed between the results of 
PRISM theory obtained
with the exact and the Lorentzian form factor are reassuring, since the 
latter has also been used for the thread model of polymers, 
and leads to the prediction that the equation of state of 
thin, non-intersecting polymers of radius of gyration $R_{\rm g}$ 
would be the same as that of infinitely thin platelets of 
radius $R=R_{\rm g}$. 

We are considering a number of extensions of the present work. 
In order to investigate the isotropic-to-nematic phase transition 
and other possible transitions to anisotropic phases, RISM and PRISM 
theories must be generalized to allow for anisotropic form factors, 
mirroring the orientation of platelets relative to a director. 
The present work will also be extended to binary mixtures of 
platelets of different diameters (and hence numbers of sites) 
and aspect ratios. This will introduce a competition between 
depletion-induced demixing and orientational ordering, which 
is expected to lead to rich phase 
behaviour~\cite{wens:04,harn:02,wens:01}.
A final extension of our work will be to charged platelets, 
modelled by using interaction sites which carry charges [19]. 
This extension is important for a description of aqueous dispersions of 
natural, or synthetic
clays, like the widely studied Laponite, which exhibit 
poorly understood Coulomb-induced gelling behaviour~\cite{mour:95}.
Our preliminary investigations~\cite{harn:02} encourage us to 
pursue a more systematic investigation of charged platelets 
within the framework of RISM and PRISM theory.

\section*{ACKNOWLEDGEMENTS}
DC wishes to thank Prof. C.~Caccamo for his continuous 
encouragement and support during this work. Helpful
discussions with Dr. R.~Blaak are also gratefully acknowledged.

\end{document}